\documentclass[american,aps,pra,reprint]{revtex4-1}

\usepackage[unicode=true,pdfusetitle, bookmarks=true,bookmarksnumbered=false,bookmarksopen=false, breaklinks=false,pdfborder={0 0 0},backref=false,colorlinks=false] {hyperref}
\hypersetup{ colorlinks,linkcolor=myurlcolor,citecolor=myurlcolor,urlcolor=myurlcolor}
\usepackage{graphics,epstopdf,graphicx, amsthm, amsmath, amssymb, times, braket, colortbl, soul, color, bm,xcolor, multirow, booktabs}
\usepackage[up]{subfigure}
\usepackage{cleveref}

\definecolor{myurlcolor}{rgb}{0,0,0.7}

\theoremstyle{plain}

\providecommand{\theoremname}{Theorem}

\newcommand*{\myproofname}{Proof}

\makeatother

\begin{document}



\title{Inhibition of spreading in quantum random walks due to quenched Poisson-distributed disorder}

\author{Sreetama Das, Shiladitya Mal, Aditi Sen(De), Ujjwal Sen}
\affiliation{Harish-Chandra Research Institute, HBNI, Chhatnag Road, Jhunsi, Allahabad 211 019, India}




\begin{abstract}
We consider a quantum particle (walker) on a line who coherently chooses
to jump to the left or right depending on the result of toss of a quantum
coin. The lengths of the jumps are considered to be independent and identically
distributed quenched Poisson random variables. We find that the spread of the walker is
significantly inhibited, whereby it resides in
 the near-origin region, with respect to the case when there is no
disorder. The scaling exponent of the quenched-averaged dispersion of the walker
is sub-ballistic but super-diffusive. We
also show that the features are universal to a class of sub- and
super-Poissonian distributed quenched randomized jumps.
\end{abstract}

\maketitle

\section{Introduction}

In the past few decades, research in quantum information science has provided a leading component towards the advancement of communication and computational technologies. This outstanding progress is due to the superiority of quantum-enabled devices over their classical counterparts. Examples include quantum dense coding \cite{dc}, quantum teleportation \cite{tprt} and quantum key distribution \cite{key}. Here we study quantum random walks (QRWs), whose classical counterparts -- classical random walks (CRWs) -- have already been established as useful tools in classical randomized algorithms \cite{crw}. The Markov chain model of CRW has succeeded in estimating the volume of a convex body \cite{cnvx} and Markov chain Monte Carlo simulation has been able to approximate the permanent of a matrix \cite{perm}. 
Random walk in a quantum-mechanical scenario was introduced in 1993 by Aharonov \emph{et al.} \cite{aharonov}.
Since their work, QRWs have been studied extensively, using both the discrete time (\cite{meyer1,meyer2}) as well as continuous time (\cite{fari,ctw}) models. 

An important feature of a QRW that makes it so different from a CRW is the faster propagation of the wave function compared to a classical walker. This happens due to the interference between different possible paths that the wave function can propagate in. For a CRW on a line, the standard deviation goes as the square root of the number of iterations, whereas for QRWs on a line, the distribution spreads linearly (ballistic propagation) with increasing number of iterations \cite{ambi}. This trait of QRWs has been extremely helpful in developing numerous quantum algorithms, e.g. in investigating the exponentially faster hitting time of QRW over CRW \cite{ctw, grover, deotto}, and in various quantum search algorithms \cite{page, kempe}. On the other hand, in the field of condensed matter, there has been investigations to realize topological phases 
that are not possible to be described by local order parameters,
in controlled systems   composed of photons \cite{otterbach} or cold gases in optical lattices \cite{sorensen, osterloh}. 
In \cite{etopo}, it has been shown that discrete time QRWs permit the experimental study of the whole class of topological phases  in one and two dimensions \cite{schnyder, kitaev}.

One-dimensional QRWs have been experimentally realized in a number of physical systems e.g. in trapped ions \cite{TA1,TA2}, nuclear magnetic resonance systems \cite{NMR1,NMR2,NMR3}, photons in waveguides \cite{WG1,WG2, WG3}, to mention a few. QRWs have  also been  applied in simulation of physical processes like photosynthesis \cite{sension, mohseni}, quantum diffusion \cite{godoy},  and breakdown of electric-field driven systems \cite{oka, oka2}. See Refs. \cite{watrs2, vazi, algo1, childs, cmqw2, exloc, bach}
for further applications.

However, this ballistic propagation of the  wave function is significantly inhibited when randomness is introduced in the substrate or medium \cite{exloc,keating,lavicka}. More precisely, an inhomogeneity in the medium breaks the periodicity of the medium and hence gives rise to suppression of spread of the wave function at certain regions/points of the lattice which remains unchanged with time. This is similar to 
the 
localization phenomena in condensed matter physics, first studied by Anderson \cite{anderson} in the context of electron localization in a disordered lattice. Another way of obtaining such reduction in spread 
in QRWs is by introducing disorder in the operations that control the dynamics of the system instead of directly making the medium inhomogeneous. This kind of inhibition of spread in discrete-time QRWs is observed by inducing disorder in the coin rotation at each iteration or by introducing phase-defects at selective sites. In the first case, during each coin flip, the rotation angle of the coin is randomly selected from some probability distribution \cite{cm1, edge}. In the second case, the quantum walker picks up a particular phase factor whenever it passes through some particular site or sites \cite{li, zhang}. 


In this work, we focus on discrete-time QRWs with a different type of channel of disorder. We introduce a quenched Poisson-distributed randomness in the length of the jump that the quantum walker takes after each coin toss, and study the resultant probability distribution on the position space after a large number of iterations. Poisson distributions with different means are considered. We observe that the walker is constrained to remain near its initial position, with the quenched averaged spread being in a regime that is 
sub-ballistic but super-diffusive. The qualitative behavior of inhibition of spread is independent of the mean. We also find that the feature remains qualitatively unaltered in 
systems where the jumps have certain sub- and super-Poissonian distributions. 
We have also studied the differences in the response on the quenched averaged spread by changing the disorder from dynamic to static.
Disorder in the jump of the walker can potentially be realized in systems where QRWs have been studied experimentally. For example, in the QRW of a single laser-cooled Cs atom 
on a one-dimensional optical lattice \cite{karski}, errors in the voltage that controls the movement of the atom from one lattice site to another during the shift operation can be modelled by a QRW with a disorder in its jump. A similar possibility exists for QRWs executed using 
$^{25}$Mg$^{+}$ ions on a lattice
\cite{TA2}. See also \cite{eckert}. 

The paper is organized as follows. In the next section, we give a short introduction to discrete-time quantum walks on a line. In Section \ref{section-tin}, we briefly describe the concept of quenched disorder and the corresponding quenched averaging, while in Section \ref{section-char}, we formally define the Poisson distribution. In Section \ref{section-panch}, we present our results on the effect on a QRW of a Poisson-distributed quenched random variable being used as the length of the jump of the quantum random walker. 
In Section \ref{section-chhoi}, we consider the case when the Poisson distribution is replaced by certain sub- and super-Poissonian distributions. Section \ref{section-aat} considers the case of static quenched disorder. We present a summary in Section \ref{section-sath}.

\section{Discrete-time quantum walk}
\label{section-dui}
In analogy to CRWs, the displacement of the particle on the one-dimensional lattice in discrete-time quantum walk is associated with the tossing of a ``quantum coin". Suppose that $\mathcal{H}_p$ denotes the Hilbert space corresponding to the position of the particle. For a one-dimensional walk of \(T\) ``iterations'', a basis of \(\mathcal{H}_p\) is $\{|i\rangle : i \in [-T, T] \cap \mathcal{Z}\}$, with \(\mathcal{Z}\) being the 
set of all integers. The Hilbert space, $\mathcal{H}_c$, of the coin is spanned by two basis states, say, $|0\rangle, |1\rangle$. But unlike CRWs, the state of the quantum coin can be in superposition of the two basis states. The particle executing the quantum random walk moves one step towards right if the coin state is $|0\rangle$ and towards left if the coin state is $|1\rangle$, but unlike CRWs, the process happens coherently, much like the quantum parallelism in quantum computer circuits \cite{qc-boi}. This conditional shift operation is described by the operator
\begin{eqnarray}
\label{shift}
\tilde{S}=\sum_{i=-T+1}^{T-1}\left(|0\rangle\langle 0|\otimes|i+1\rangle\langle i| + |1\rangle\langle 1|\otimes
|i-1\rangle\langle i|\right).\quad
\end{eqnarray}
The random walk procedure begins with a rotation in the coin space, which is analogous to the tossing of a coin in CRW.
The coin rotation can be any unitary operation on the coin Hilbert space, thus generating a rich family of random walks. Here we consider the Hadamard coin for which the initial rotation is the Hadamard gate, 
given by
\begin{eqnarray}
\label{unitary}
H=\frac{1}{\sqrt{2}}\left(\begin{array}{cc}
1 & 1  \\ 
1 & -1  \\ 
\end{array}\right),
\end{eqnarray}
Suppose also that initially the particle is at the origin, for which the particle state is $ |0\rangle $, and that the initial state of the coin is $ |0\rangle $. In each iteration of a given run of the experiment, we apply the Hadamard rotation on the coin and then apply the shift operation on the joint system of the coin and the particle. So, after the first iteration, the joint state of the coin-particle system can be 
represented as 
\begin{eqnarray}
\label{total}
\tilde{S}(H\otimes\mathbb{I})|0\rangle\otimes 0\rangle =\tilde{S}\frac{1}{\sqrt{2}}(|0\rangle +|1\rangle)\otimes |0\rangle\nonumber\\
=\frac{1}{\sqrt{2}}(|0\rangle\otimes |1\rangle +|1\rangle\otimes |-1\rangle),
\end{eqnarray}
where $\mathbb{I}$ denotes the identity operator on \(\mathcal{H}_p\).
We iterate this process \(T\) times without performing any measurement at the intermediate iteration times. Therefore, after $ T $ iterations, the state of the coin-particle duo reads $[\tilde{S}(H\otimes\mathbb{I})]^T |0\rangle \otimes |0\rangle $. For a CRW, in the limit of a large number of iterations, the position of the particle is Gaussian distributed, with the standard deviation diverging only as $ \sqrt{T} $. On the other hand, the state $ [\tilde{S}(H \otimes \mathbb{I})]^{T} |0\rangle |0\rangle $ has a standard deviation that diverges as $ T $. Note that in the limit  $ T \rightarrow \infty $, the scaling of standard deviation of the probability distribution of the walker with respect to number of steps taken, was analytically derived in \cite{konno, grimmett, konno1}.

\section{Quenched disorder}
\label{section-tin}

\begin{figure}
\includegraphics[scale=0.5]{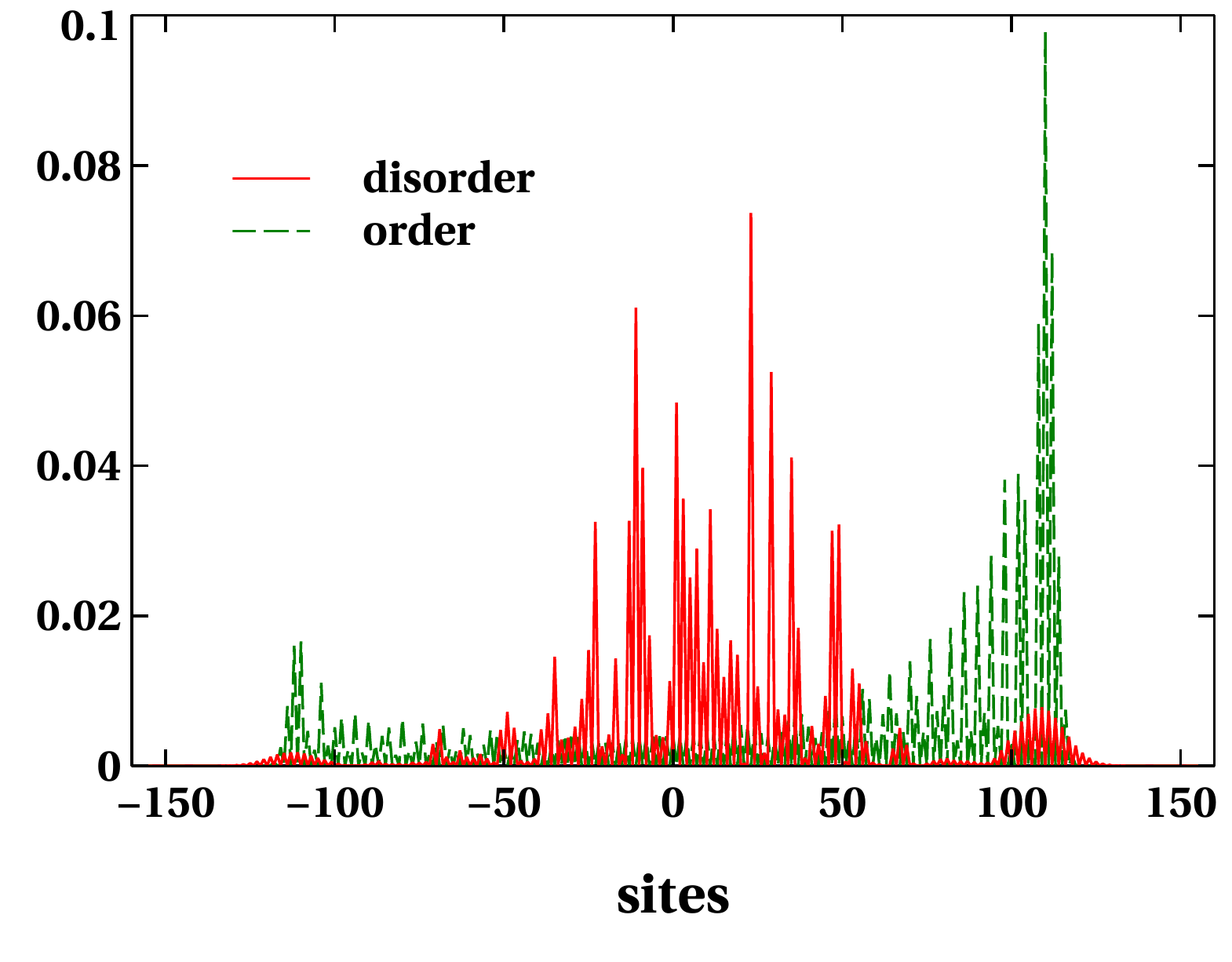}
\caption{Comparison of the ordered quantum walker with a disordered one. 
We compare the site probabilities of a quantum random walker without disorder (green, dashed) with the same for one with 
Poisson-distributed disordered step-size (red, solid), for 160 iterations.
The vertical axis represents the site probabilities after 160 iterations, while the horizontal axis represents the 
sites, with the origin of the horizontal axis representing the initial position of the walker. 
The probabilities for the disordered case are for a particular realization of the disorder.
A comparison of the site probabilities for the disordered walk, with those of the ordered case 
clearly indicates that disorder 
causes the quantum walker to remain in the near-origin region, albeit only for the particular realization of the disorder. The same feature is seen for other realizations of the disorder, and we will see in subsequent analysis that a quenched averaging over a large number of disorder realizations provides a clearer signature of what is akin to ``localization'', and which is referred to in this paper as 
``inhibition of spread''. Both axes represent dimensionless quantities.}
\label{distri}
\end{figure}


In this work, we consider  a category of disorder, which, to our knowledge, has not yet been studied for quantum random walks. In a QRW, at every iteration, the walker is displaced by one step, conditioned on the coin state. There is, therefore, an already existing randomness in quantum walks due to the superposition between the two coin states that is created at each iteration by the Hadamard gate. Let us introduce an additional randomness in the amount of displacement of the particle at each time step. In \cite{lavicka}, Lavi{\v c}ka \emph{et al.} introduced a randomness in the jump length for quantum random walk in optical multiports, where they considered that the quantum walker, at each coin toss, can jump to the next multi-port with some probability $ \delta $, or can connect to a multiport at a fixed distance with probabilty $ (1-\delta) $. Unlike Lavi{\v c}ka \emph{et al.} and unlike in the case without disorder, 
in our case, after each coin toss, 
the walker can jump an arbitrary number of steps with the length, \(j\), of the jump being randomly distributed according to a certain discrete probability distribution \({\cal P}_{{\cal R}}(j)\), where 
\({\cal R}\) denotes the effective maximal jump. This jump length $ j $ is the same irrespective of which vertex the particle is in at that time-step.
Moreover, the jump length is also the same, albeit in different directions, regardless of whether the quantum coin is ``thrown'' into the $|0\rangle$ or the  $|1\rangle$ state by the Hadamard operator of that iteration. Note that, when $ j = 0 $, the walker stays at its current position.
  Introduction of this kind of disorder can be described by the shift operator given by
\begin{eqnarray}
\label{anami}
\tilde{S}^{\prime}=\sum_{i=-(T-1){\cal R}}^{(T-1){\cal R}}\left(|0\rangle\langle 0|\otimes|i+j\rangle\langle i| + |1\rangle\langle 1|\otimes|i-j\rangle\langle i|\right),\nonumber \\
\end{eqnarray}
where \(j\) takes values from $\{0, 1, \ldots, {\cal R}\}$ according to the distribution \({\cal P}_{{\cal R}}(j)\), and  
the coin operation is taken to be Hadamard. 
Values of $ j $ higher than \({\cal R}\) are either non-existent or are ignored for some physical reason (e.g. insignificant effect on the position probabilities for allowing \(j>{\cal R}\)).
Note that the Hilbert space of the walker has now changed into one that is spanned by 
\(\{|i\rangle:i\in [-T{\cal R}, T{\cal R}] \cap {\cal Z}\}\).

The disorder that is introduced in the step length at every iteration of the protocol is ``quenched", so that it remains fixed for the entire span of a particular run of the protocol. To obtain a meaningful value of a physical quantity, say, the dispersion of a walker in a quenched disordered system, one must perform a configurational averaging over the disordered parameters. Note that this averaging needs to be performed only after all other calculations have already been performed. Such an averaging is referred to as ``quenched averaging". We are in particular interested in quenched averaged dispersion of QRWs, in which the step length at different iterations of the protocol are independent and identically distributed quenched  random variables distributed as \({\cal P}_{{\cal R}}(\cdot)\).

\section{Poisson distribution}
\label{section-char}

The Poisson distribution, due to A. de Moivre and S. D. Poisson, is a discrete probability distribution which gives the probability of the number of occurences of a certain event in a fixed interval, as 
\begin{equation}
\label{poisson}
p(k) = \dfrac{e^{-\lambda} \lambda^{k}}{k!},
\end{equation}
where $ \lambda $ is the average number of events that occur in the given interval and $ p(k) $ is the probability that the event will occur $ k $ times in that interval. The Poisson distribution is known to be useful in a large variety of situations.
Examples include the number of mails received per day by a particular office, the number of scientific papers published in a month from a certain institute, the number of trains canceled in a week on a particular route, etc. In this work, we begin by using the Poisson distribution around the average $ \lambda= $ 1 to randomly generate the integer values of $ j $ (see Eq. (\ref{anami})). For numerical convenience, we have discarded all those random outcomes where $ j > $ 5, and have renormalized the resulting distribution. Note that for $ \lambda= $ 1, the probability that $ j>5 $ is of the order of $ 10^{-4} $.

\section{Inhibition of spread}
\label{section-panch}
\begin{figure}
\includegraphics[scale=0.45, angle=0]{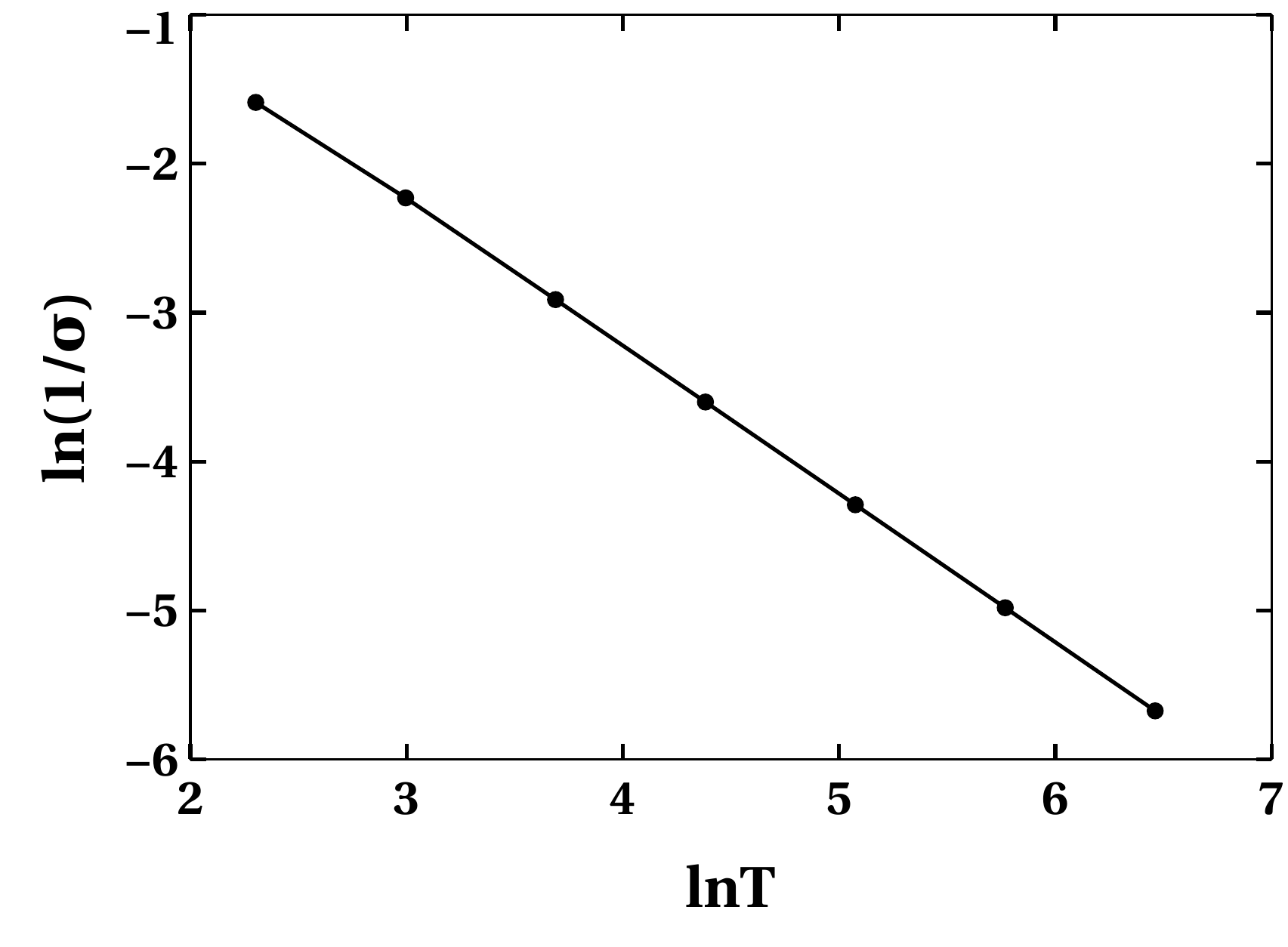}

\caption{Scaling behavior of the dispersion in the ordered qauntum random walk. We plot $ \ln(1/\sigma) $ against $ \ln T $ for up to 640 iterations to find a straight-line fit, and with a slope of \(-1\). All quantities are dimensionless. See \cite{konno, grimmett,konno1}.}
\label{fig1}
\end{figure}

Let us first briefly examine the results for the discrete quantum walk with no disorder in the system. We assume the coin to be initially in the state $ |0\rangle $ and the particle to be initially at the origin. We apply the Hadamard gate on the coin, following which the shift operator as in Eq. (\ref{shift}) is applied on the particle. This process is repeated several times, and the probability distribution of the walker's position after 160 iterations is depicted in Fig. \ref{distri}. We find that the walker has a high probability to be around \(i=100\) after 160 iterations. 
We are mainly interested in 
studying the standard deviation of the probability distribution in the particle space. In this case, as expected, the standard deviation, $ \sigma $, varies linearly with the number of iterations, $ T $. We perform a log-log scaling analysis between $\ln(1/\sigma) $ and $\ln T $ (see Fig. \ref{fig1}) to find a straight-line fit. The slope of the straight line is tan$(-\pi/4)=-1$. A QRW with a  $ \sigma $ that is linearly varying with respect to the number of iterations is usually referred to as ``ballistic propagation'' of the particle.

QRWs appears in several colors and hues, encompassing discrete as well as continuous walks. Disorder in such systems have also been incorporated in different ways. This includes e.g. \cite{keating}, which incorporates an imperfection in the graph which supports a continuous-time walker, resulting in an inhibition of spread of the latter on the graph at the starting point. Another work \cite{exloc} associates a transgression in the dynamical equation of the continuous-time quantum random walker, wherein there can appear situations 
where
the walker remains virtually unmoved. The corresponding reduction in spread depends on the type of disorder involved, and the consequences can also vary from being ``diffusive'' (standard deviation of the walker is proportional to the square root of the number of iterations) to being ballistic. A discrete-time QRW with non-Hadamard operations at each toss of the quantum coin was considered in \cite{cm1}. The non-Hadamard operator was chosen to be different at each iteration, and the result was a 
suppression 
of the wave function of the walker to its initial point. Further such cases can be found e.g. in \cite{joye,ahlbrecht}.
%
In another example, Ref. \cite{zhang} finds that a non-Hadamard quantum coin associated with a discrete-time quantum walker can confine or repulse the walker at or from its initial point depending on the phase of the rotation in the quantum coin at each iteration. See also \cite{edge,li} in this regard.

\begin{figure}
\includegraphics[scale=0.45, angle=0]{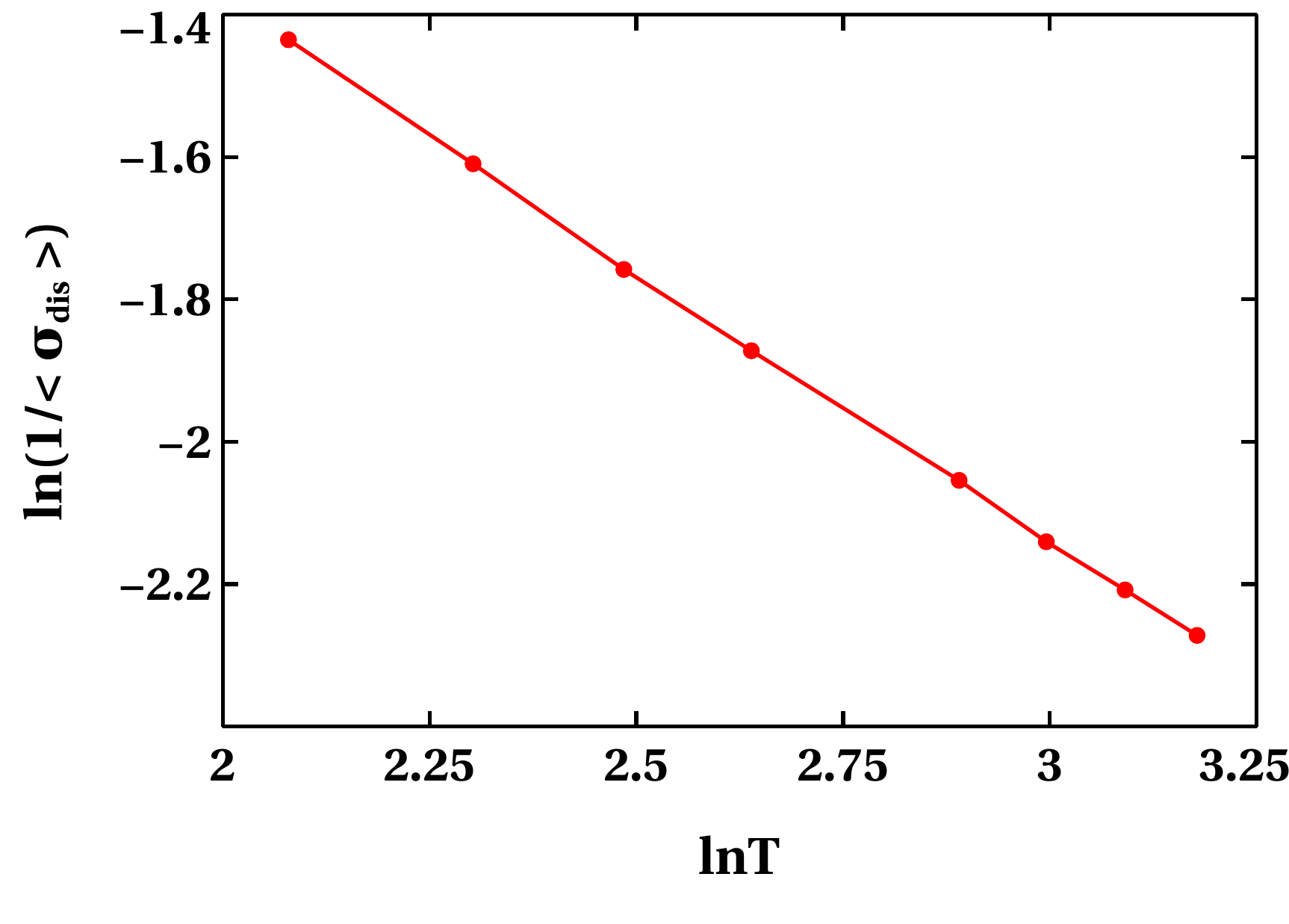}
\caption{Scaling behavior of dispersion in the quenched-disordered quantum random walk. We plot $ \ln(1/\langle\sigma_{dis}\rangle) $ against $ \ln T $ for up to 24 iterations to  find a straight-line fit, just as in Fig. \ref{fig1}. However, the slope in the disordered case is given by $\tan\theta\approx -0.8$. All quantities are dimensionless.}
\label{fig2}
\end{figure}

In this work, we consider the quantum random walk, where we have included a disorder in the number of steps, $ j $, the particle can go after each tossing of the coin. We begin by
examining the case where $ j $ is randomly chosen from the Poisson distribution with unit mean. 
Here we first apply the Hadamard gate on the coin, following which the shift operator as in Eq. (\ref{anami}), is applied on the particle. After $ T $ iterations, we calculate the standard deviation $ \sigma_{dis} $, for the particular realization of the disordered variables. 
In Fig. \ref{distri}, we provide a comparison between the probabilties in the cases when there is no disorder, and when 
there is a particular realization of the Poisson-distributed disorder. 
It is clear from the figure that the disorder hinders the walker's movement to regions 
away from its initial position, albeit only for the particular realization of the disorder considered in 
Fig. \ref{distri} (cf. \cite{keating, exloc}). We will see that the inhibition in spread observed here persists even after a quenched  averaging.
The jump or shift, $ j $, at any iteration of a particular run is considered to be independent from but identically distributed with the jump in any other iteration of that run. The physically relevant quantity, however, is the average of this $ \sigma_{dis} $ for different realizations of the disorderd variables. We denote this quenched avaraged $ \sigma_{dis} $ as $ \langle\sigma_{dis}\rangle $. Our numerical simulations show that with increasing number of iterations, $ T $, $ \langle \sigma_{dis} \rangle$ diverges to infinity, just like in case of the ordered system, but the divergence is much slower. The quenched averaging is performed over 4000 disorder realizations. 
Here, the ``finite-size" scaling exponent is $\approx 0.8$, as compared to unity in the case of the ordered system. The ``finite-size" here refers to the finite number of iterations, and corresponds to the finite number of subsystems in finite-size scaling analysis in many-body physics. 
The finite-size analysis can be stated more precisely by expressing the disorder-averaged dispersion as
\begin{equation}
\ln\left(\dfrac{1}{\langle\sigma_{dis}\rangle}\right) = -\alpha \ln T + \ln A,
\end{equation}
so that
\begin{equation}
\langle\sigma_{dis}\rangle = A^{-1}T^{\alpha},
\end{equation}
where $ A\approx 1$ and $ \alpha \approx 0.8 $, up to 
the first significant figure. See Fig. \ref{fig2}, and compare with Fig. \ref{fig1}.
The disorder, therefore, induces a standard deviation of the walker that is intermediate to being ballistic and 
diffusive. It is sub-ballistic but super-diffusive. 

We also try to look at the effects of changing the mean of the Poisson distribution to values other than unity. The results have been summarized in Table \ref{table2}. We observe that for different means, the value of the scaling exponent is in the range $-0.8$ to $-0.7$. 

\begin{table}[]
\renewcommand{\arraystretch}{1.5}
\begin{tabular}{|c|c|c|}
\hline
\textbf{Distribution}    & \multicolumn{1}{l|}{\textbf{Mean}} & \multicolumn{1}{l|}{\textbf{\begin{tabular}[c]{@{}l@{}}Scaling\\ exponent\end{tabular}}} \\ \hline
\multirow{4}{*}{Poisson} & 0.5                                & -0.8                                                                                     \\ \cline{2-3} 
                         & 1.0                                & -0.8                                                                                     \\ \cline{2-3} 
                         & 1.5                                & -0.7                                                                                     \\ \cline{2-3} 
                         & 2                                  & -0.7                                                                                     \\ \hline
\end{tabular}
\caption{Sub-ballistic but super-diffusive spread for quenched Poisson disorder with different values of the mean.
The tabular data presents the scaling exponent $ \alpha $ when the jump length of the quantum walker is chosen from Poisson distributions having different mean values.}
\label{table2}
\end{table}

Below we find that this behavior of having an intermediate scaling exponent (sub-ballistic but super-diffusive) of standard deviation is 
far more general, and can be seen in types of disorder, widely varying from the Poissonian one.

\begin{table}[]
\renewcommand{\arraystretch}{1.5}
\begin{tabular}{|c|c|c|c|}
\hline
\textbf{Class}             & \textbf{Distribution}                      & \multicolumn{1}{l|}{\textbf{Variance}} & \multicolumn{1}{l|}{\textbf{\begin{tabular}[c]{@{}l@{}}Scaling\\ exponent\end{tabular}}} \\ \hline
\phantom{Poisson}                           &  Poisson                                  & 1                                      & -0.8                                                                                    \\ \hline 
\multirow{3}{*}{Sub-Poissonian}   & \multirow{2}{*}{Binomial}          & 1/2                                    & -0.8                                                                                    \\ \cline{3-4} 
                                  &                                    & 8/9                                    & -0.8                                                                                    \\ \cline{2-4} 
                                  & Hypergeometric                     & 1/3                                    & -0.8                                                                                    \\ \hline
\multirow{3}{*}{Super-Poissonian} & \multirow{2}{*}{Negative binomial} & 2                                      & -0.8                                                                                    \\ \cline{3-4} 
                                  &                                    & 10/9                                   & -0.7                                                                                    \\ \cline{2-4} 
                                  & Geometric                          & 2                                      & -0.8                                                                                    \\ \hline
\end{tabular}
\caption{Sub-ballistic but super-diffusive spread for different classes of quenched disordered discrete distributions of the jump length in a quantum random walk. We present here 
the values of the scaling exponent $ \alpha $ obtained in cases when the jump length is randomly chosen from Poisson and certain paradigmatic sub- and super-Poissonian distributions. The corresponding  variances are also indicated in the table. All the distributions have unit mean.}
\label{table1}
\end{table}


\section{Sub- and Super-Poissonian distributions}
\label{section-chhoi}

From
the Poisson distribution, let us now move over to 
one-dimensional QRWs where $ j $ is randomly chosen according to certain paradigmatic sub- and super-Poissonian distributions. A sub- (super-) Poissonian distribution has a smaller (larger) variance than the Poisson distribution having the same mean. As examples of sub-Poissonian distributions, we consider the binomial and hypergeometric distributions, while as examples of super-Poissonaian distributions, we 
perform our analysis by considering the negative binomial and geometric distributions.

The \emph{binomial distribution}  is a discrete probability distribution involving Bernoulli trials, with the latter being independent and identically distributed (i.i.d.) trials that have two outcomes, called ``success'' and ``failure''. The total number of trials is fixed to a certain integer \(n\), and the 
random variable is the number of successes, \(k\), occuring at each trial with probability $ p $. The probability mass function (pmf)  is given by $ \binom{n}{k}p^{k}(1-p)^{n-k} $, with the mean and variance being  $ np $ and $np(1-p) $ respectively.
Note that the variance is lower or equal to the mean.

The \emph{hypergeometric distribution} is a discrete probability distribution where one is given a finite population having size $ N $ within which  there are exactly $ K $ elements that we refer to as ``successes''. The random variable is the number \(k\) of successes in a particular trial of $ n $ draws without replacement. 
The pmf of the hypergeometric distribution is therefore given by $ \frac{\binom{K}{k}\binom{N-K}{n-k}}{\binom{N}{n}} $. Unlike the binomial distribution, here, after each draw, the probability of success changes. The mean and variance of this distribution is given by $ \frac{nK}{N} $ and $ \frac{nK}{N}\frac{N-K}{N}\frac{N-n}{N-1} $, so it has a varinace lower than the mean, i.e. it is a sub-Poissonian distribution.

The \emph{negative binomial distribution} is also a discrete probability distribution involving Bernoulli trials. The random variable in this case is the number of successes, \(k\), 
until a specified number, \(r\), of failures, with the fixed probability of success in each trial being \(p\). The corresponding pmf is given by $ \binom{k+r-1}{n}(1-p)^{r}p^{k} $.
The mean of the negative binomial distribution is $ \dfrac{pr}{1-p} $, while the variance is $ \dfrac{pr}{(1-p)^{2}} $. Note that the variance is then always larger or equal to the mean.

The \emph{geometric distribution} is a discrete probability distribution involving Bernoulli trials. Here, the random variable is the number, $ k $, of failures before the first success occurs, with probability of success in each trial being $ p $, so that  the pmf is given by $ p(1-p)^{k} $. The mean and variance are given by $ \frac{1-p}{p} $ and $ \frac{1-p}{p^{2}} $, so that the distribution has a variance greater than the mean, i.e. it is a super-Poissonian distribution.

For demonstration, 
we  choose two (sub-Poissonian) binomial distributions having two different variances, respectively $ \frac{1}{2} $ and $\frac{8}{9}$, and one (sub-Poissonian) hypergeometric distribution with variance $ \frac{1}{3} $; note that the variances are \emph{smaller} than their common unit mean. Parallely, we  choose two (super-Poissonian) negative binomial distributions having two different variances, respectively 2 and $ \frac{10}{9} $, and one (super-Poissonian) geometric distribution
with variance 2; note that the variances are \emph{larger} than their common unit mean. 
For each of the six cases, we perform the scaling analysis by plotting $ \ln(1/\langle\sigma_{dis}\rangle)$ against $\ln T $. Interestingly, in all the cases, the scaling exponent reduces compared to the unit value in the ordered walk. 
The scaling exponents remain in the range $-0.8\) to \(-0.7$. 
The set of data thus obtained is summarized in Table \ref{table1}.
In all the cases, the disorder averagings are performed over 4000 realizations, and the 
effective maximal jump (\({\cal R}\)) is chosen so that the total probability to jump further is 
of the order \(10^{-4}\) or less.

\section{Effect of static disorder}
\label{section-aat}
The disorder considered in this work till now is dynamic in nature, i.e. the random jump length is different for different time-steps. We now study the effect of introducing a static quenched disorder in the jump length. In this case, associated to each site, there is a particular integer, fixed for all time, but random with respect to sites. The quantum walker, after reaching a particular site, will take the next jump having a jump length equal to the integer associated to that site. We choose the integers randomly from a Poisson distribution with unit mean, and calculate the standard deviation of the probability distribution after a certain number of steps. Then we take a large number of such random integer configurations, to find the quenched averaged  standard deviation.
We find that the quenched averaged standard deviation 
grows almost linerarly with $ T $ for a relatively small number of steps, \(T\), but for $ T>10 $, 
 it saturates to a value $ \langle\sigma_{dis}\rangle|_{T\gtrsim10} \approx 1.8 $.

\section{Conclusion}
\label{section-sath}

%
We introduced a quenched disorder in the number of steps that the quantum particle (walker) can jump after each coin toss in a discrete quantum random walk in one dimension. 
We first considered the case where the length of the jump is randomly chosen from the Poisson distribution around unit mean. We found that the spread of the walker, as quantified by its 
standard deviation,
after quenched averaging over a large number of configurations of the disorder, 
has a finite-size scaling exponent which is approximately \(20\%\) lower than that for the 
ordered case, thereby implying
a slowdown of the walker. The walker is consequently
 sub-ballistic but super-diffusive.
We then argued that this feature of the scaling exponent is generic, as it was  
found to be shared by random distributions 
widely varying from the Poissonian one. In particular, it exists in both sub- and super-Poissonian random distributions. We also performed the analysis, obtaining qualitatively similar results, for Poisson distributed quenched disorders with non-unit means. Inhibition of spread of the quantum random walker was also found for static quenched disorder. 
The effects studied can potentially be observed with currently available technology in systems where quantum random walks have been experimentally realized, in particular with atoms hopping on an optical lattice.



\end{document}